\documentclass[showpacs, preprintnumbers, nofootinbib, aps, prd, superscriptaddress,10pt, showkeys, notitlepage, twocolumn]{revtex4-1}

\usepackage{natbib}

\usepackage{amsfonts}
\usepackage{amsmath}
\usepackage{amssymb}
\usepackage{bm}
\usepackage{dcolumn}
\usepackage{graphicx}
\usepackage{graphics}
\usepackage[latin1]{inputenc}
\usepackage{latexsym}
\usepackage{rotating}
\usepackage{hyperref}
\usepackage[all]{hypcap} 
\usepackage{xspace} 
\usepackage[usenames]{color}
\usepackage{mathrsfs}

\newcommand\be{\begin{equation}}
\newcommand\bea{\begin{eqnarray}}
\newcommand\ee{\end{equation}}
\newcommand\eea{\end{eqnarray}}
\newcommand\bw{\begin{widetext}}
\newcommand\ew{\end{widetext}}

\newcommand{\p}{\partial}

\begin{document}
\title{Modelling Gaia photometry signals of dark halos}

\author{George Pappas}
\email{gpappas@auth.gr}
\affiliation{Department of Physics, Aristotle University of Thessaloniki, Thessaloniki 54124, Greece}

\author{Ippocratis D. Saltas}
\email{saltas@fzu.cz}
\affiliation{CEICO, Institute of Physics of the Czech Academy of Sciences, Na Slovance 2, 182 21 Praha 8, Czech Republic}

\author{Laurent Eyer}
\email{laurent.eyer@unige.ch}
\affiliation{Department of Astronomy, University of Geneva, Chemin Pegasi 51, 1290 Versoix, Switzerland}

\begin{abstract}
We use the framework of microlensing to show that observations of binary systems, such as those made by {\it Gaia}, combined with follow-up weak lensing measurements, can provide a means to probe halos of exotic matter, possibly clumped around compact objects such as black holes. This could potentially cover a broad range of physical scenarios - from dark matter mini-halos to black holes with bosonic configurations around them, known as hair. In a dark halo-regular star binary system, assuming that light can freely propagate through the halo of the exotic matter, the companion star will produce characteristic, sizable lensing signatures due to the lens's potential deviating from a point mass gravitational potential. The signature of the multiple images, the magnification and the light-curve could be the smoking gun of such a dark halo structure. We discuss how the precise observations of the {\it Gaia} survey offer an opportunity to search for new, yet undiscovered fundamental fields interacting gravitationally with baryons.
\end{abstract}


\pacs{}

\date{\today}
\maketitle


\section{Introduction}
\label{sec:Intro}
We are living in an exciting time for astrophysics. Current high-precision surveys for astrophysics and astronomy such as {\it Gaia} or {\it Euclid} \footnote{https://sci.esa.int/web/gaia, \\ 
https://sci.esa.int/web/euclid} offer an unprecedented window into the physics of astrophysical systems. From a fundamental physics standpoint, such surveys promise to shed light on fundamental questions of modern physics, such as the nature of dark matter and dark energy, which, in turn, ultimately relates to our understanding of particles and forces in the Universe.

Indeed, a new pressure-less (dust-like) component of matter in the Standard Model of particle physics is required by a plethora of observational probes at larger or smaller scales in the Universe, e.g. the Cosmic Microwave Background, galactic rotation curves, or supernovae observations \citep[see e.g.][]{Famaey:2011kh, Hui:2016ltb, Roszkowski:2017nbc}. 
 At the same time, theories beyond the Standard Model typically predict the existence of new particles of various masses and spins. Within an astrophysical context, this begs the question of whether stable configurations of such exotic matter can form, be it dark matter or other fundamental fields. Clumps of exotic matter could form either isolated or around compact objects such as black holes (BHs), through either the process of gravitational collapse  \citep{Gondolo_Silk1999,FranzSchunck_2003,Sadeghian:2013laa,Brito:2015pxa,Pombo:2023ody}, or the so--called mechanism of scalarisation/vectorisation \citep{Cardoso:2013fwa,Silva:2017uqg,Dima:2020yac,Cunha:2019dwb,Dolan:2018dqv,Pani:2012vp}. The latter intimately relates to the question of the nature of BHs, and the existence of hair-like configurations around them \citep{Liebling_2023, Herdeiro_etal2021,Berti_etal2021}.  Our proposed probe will set out a promising and powerful tool to detect exotic structures in our galaxy, as well as place constraints on the parameters of their profiles from observations.

In this work, we introduce a framework to probe the existence of stable configurations of clumps of exotic matter based on the combination of astrometry and lensing photometry observations of binary systems within our galaxy, where the binary system is comprised of such a dark exotic matter configuration and a regular star that is lensed by its dark companion. 
The astrometry frontier is currently led by the high-precision observations of {\it Gaia} \citep{GAIA,GAIA1,GAIA_DR3}, and promises to drastically improve in the future. Furthermore, Gaia (spectro-)photometry also provides a unique all sky multi-epoch survey.
On the other hand, several precision surveys aiming at (micro-)lensing are underway while others have already been completed  \citep{Alcock_2000,EROS-2:2006ryy,OGLE-IV}. 
Photometric self-lensing in binary systems was first predicted by Maeder \cite{1973A&A....26..215M} and observed four decades later by Kruse and Agol \cite{2014Sci...344..275K}. The periodic nature of the phenomenon not only facilitates its detection but also opens the possibility for dedicated follow-up observations, enabling detailed characterisation of the system's orbital and physical parameters, potentially including the halo properties of the black hole or degenerate star.

Dark matter models are conventionally split into two broad families, namely Weakly Interacting Massive Particles (WIMPs) and fuzzy dark matter models \citep{Roszkowski:2017nbc}. In the first case, dark matter assumes the form of a pressureless massive particle (dust particle), while in the second case, that of a dynamical field with an associated de~Broglie wavelength \citep{Hui:2016ltb}. The latter property of fuzzy models allows for a richer phenomenology and aims to address certain shortcomings of WIMPS at short scales. 

Under certain conditions, particles beyond the Standard Model of any spin, which may or may not contribute to dark matter, can form halos at astrophysical scales. This may be the result of collapse under their own gravity, or through clumping around some compact object such as a BH \footnote{The evolutionary channels forming such systems play a key role but will not be the topic of this work.}. This includes the case of the formation of dark matter structures at small (sub-galactic) scales, the bosonic condensates known as boson stars, or the fundamental question of BH physics, that is, whether BHs may develop ``hair" configurations around them made out of some exotic field. 

The key novelty of our work concerns on applying the microlensing framework on searching for dark haloes (DHs) with {\it Gaia}-like binaries (involving the relevant masses, binary sizes, etc.). Our discussion will cover all dark matter theories or scenarios where exotic matter is predicted to form clumps of matter at astrophysical scales, irrespective of the formation channel. The fine details of the profile characterising the scaling of the density of the halo with the radius is certainly a model-dependent quantity. In practice, a parametrization is enough to capture all the essential features exhibited in full-fledged numerical simulations, a notable example being the Navarro-Frenk-White profile for dark matter \citep{NFW}. Our proposed probe will set out a promising tool to place constraints on such profiles from small-scale astrophysical observations. It will complement and extend similar previous works where the lensing signatures of (isolated) exotic matter configurations were studied assuming the thin-lens approximation \citep{Croon_2020a,Croon_2020b,Croon_2024, Romao_2024}. One of the key aspects of our work is the focus on binary systems which are within the reach of the {\it Gaia} survey. Such binaries can be observed photometrically by both {\it Gaia} as well as ground-based photometry telescopes doing follow-up observations. The combination of high-precision astrometry with photometric observations offers a highly promising framework for the detection and analysis of such systems.

As we will see, for binary systems such as those recently discovered by {\it Gaia} (BH1 \cite{El-Badry2023,Chakrabarti2023AJ....166....6C}, BH2 \cite{BH2}, BH3 \cite{BH3}) the existence of the DH can leave distinct signatures in the lensing signal, within the precision of current instruments. This is predominantly imprinted on the structure of the multiple images and the resulting magnification or the light curve, respectively. Our analysis shows that the change in apparent magnitude of the star during a transit (i.e., when the star enters the ``shadow'' of its DH lens companion), depending on the geometry, can be as large as a few magnitudes during ingress/egress or a few tenths of a magnitude during the rest of the transit.
The proposed framework relies on the combination of {\it Gaia}'s astrometric measurements with follow-up lensing observations of the binary system. This synergy is key - the precise reconstruction of the orbital parameters from {\it Gaia} will not only allow an accurate follow up of the binary, but will also break degeneracies with lensing observables. 

{\it Gaia}'s astrometric precision is clearly very high, and as expected, depends on the apparent brightness of the target source. The predicted parallax uncertainty
for a stellar source of $G=15$ mag is estimated to be $22 \, \mu \rm{as}$ for DR4, and $16 \, \mu \rm{as}$ for DR5 respectively.  Notice that, astrometric precision degrades with fainter sources.
The geometric parameters of a binary system (e.g. period, parallax, proper motion) are found through a simultaneous fit of all geometrical/orbital parameters. For example, for the binary system BH3, the standard deviation of the along scan astrometric residual is $\sim 0.14\, \rm{mas}$\footnote{private communication of B.Holl.}.

As regards {\it Gaia}'s photometry, it is also exceptional. Still using the example of a source with G-band magnitude $G=15$, the photometric precision on the mean G-band equals $0.2 \, \rm{mmag}$ for DR4, and $0.1 \, \rm{mmag}$ for DR5 respectively. 

We structure the paper as follows: In Section \ref{sec:weak-lensing} we introduce the theoretical framework for weak lensing by a DH, and in Sections \ref{sec:magnification}, \ref{sec:magnification2}, and \ref{sec:Detection}, we discuss the key phenomenological imprints and population estimates of such systems. We summarise in Section \ref{sec:summary}.

\section{Weak lensing through a dark halo} \label{sec:weak-lensing}
\begin{figure}
 \centering
\includegraphics[width=0.35\textwidth]{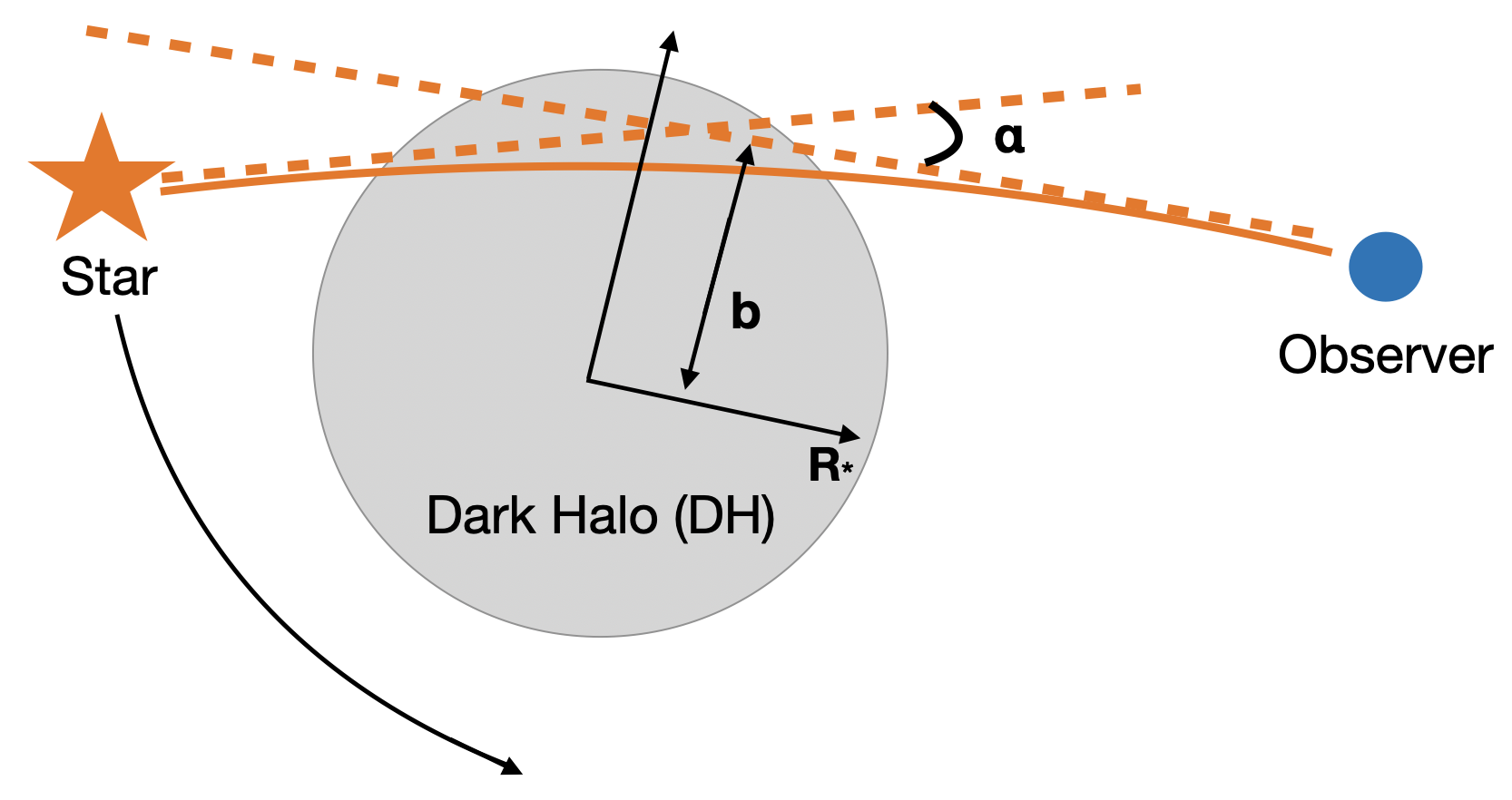}
\caption{
A schematic representation of our binary setup. The gravitational lens (DH) is a self-gravitating clump of exotic matter with or without a BH at the centre. The light source in orbit is a stellar object such as the one in {\it Gaia}'s BH1 binary system. Propagation of light through the density of the DH induces non-trivial signatures of multiple images and magnification, which can be observed through photometric observations. {\it Gaia}'s astrometry and photometry observations allow us to identify such binary systems in our galaxy and probe a broad range of theories for dark matter and other particles beyond the Standard Model. The impact parameter ($b$) and deflection angle ($\alpha$) are explained in the text.}
\label{fig:schem}
\end{figure}

Here we will introduce the main equations we will use, summarising previous results in the literature \citep{2018bookCongdon}. In the weak-field approximation, gravitational lensing can be described in terms of Fermat's principle, where spacetime has an effective index of refraction, 
$
n=\frac{c}{c'} = 1-\frac{2\Phi}{c^2},
$
where $\Phi$ satisfies the Poisson equation $\nabla^2 \Phi  = 4\pi G \rho$. By extremising the length of the path of a light-ray, $\delta \int_A^B\frac{c}{n}dt = 0$,
one arrives at an expression for the deflection angle, which is given as 
\be \vec{\alpha} = -\frac{2}{c^2}\int \vec{\nabla}_{\perp}\Phi d\lambda, \ee
where $\vec{\nabla}_{\perp}\Phi$ is the gradient of the gravitational potential perpendicular to the light-ray, while the integration takes place along the light-ray. A schematic representation of the process can be seen in Figure \ref{fig:schem}.

For a compact and point-like source of mass $M$, such as a BH, in the weak lensing region one has the usual Newtonian gravitation potential of a point source in empty space, i.e., $\Phi=-GM/r$. In the case of a DH, things would be similar as long as the light-rays propagate outside it, but there will be differences. In particular, depending on how extended the halo is, light can be propagating in regions that are inside the cloud. In that case, the gravitational potential would be modified to be
\be 
\Phi_{\textrm{DH}} \propto \frac{GM(r)}{r}, 
\ee 
where $M(r)$ will be related to the mass profile of the scalar halo. The dependence of the mass on radius, as opposed to a constant mass, can dramatically change the lensing properties of the gravitational lens, compared to what one would have in the case of a very compact companion (such as a BH) and propagation of light in a vacuum.

In the simplest possible case, where one can assume that the DH has a uniform density $\rho = \rm{const.}$, the mass distribution would be $M(r)=(4/3)\pi \rho r^3$ and the potential in terms of the total mass $M$ and the radius $R$ will be $\Phi_{\rm in}(r)=\frac{GM}{2R^3}\left(r^2-3R^2\right)$. Then, assuming a small deflection angle $\alpha \ll 1$, the deflection will be in the radial direction and with a magnitude given by 
\be \alpha = \frac{2}{c^2}\int_{-\infty}^{+\infty}\frac{\p}{\p b}\left(\frac{GM(x;b)}{\sqrt{b^2+x^2}}\right)dx, \ee
where the integration is taking place along the path of the light-ray ($x$). Here, $b$ is the impact parameter of the ray (see also Fig. \ref{fig:schem}) and $M(x;b)$ is the effective mass profile, while $r=\sqrt{x^2+b^2}$. When $b\leq R$, $M(x;b)$ is given by $M(x;b)=\frac{GM}{2R^3} \left(b^2+x^2\right)^{3/2}$, while for $b>R$ it is a constant, $M(x;b)=-GM$. We can then express the deflection angle $\alpha$ as 
\be \alpha =  \left\{  
          \begin{tabular}{cc} 
           $\frac{4 G M}{b c^2}$ & $b>R$\\  
           $\frac{4 G M}{b c^2}\left(1-\frac{a^3}{R^3} \right)$& $b\leq R$
           \end{tabular},\right.
\label{eq:deflection}
\ee
where $a^2=R^2-b^2$. 
This has also been derived by \cite{Croon_2020a} (see also \cite{Croon_2020a,Croon_2020b,Croon_2024, Romao_2024}).
We can define the characteristic quantity,
$ \alpha_o\equiv\frac{GM}{c^2 R} \approx 0.425'' \times \left(\frac{R_\odot}{R}\right)\left(\frac{M}{M_\odot}\right), $
where for a lens with the mass and radius of the Sun, we recover a deflection angle $\alpha_\odot = 4GM_\odot/c^2R_\odot \approx 1.7''$. Since neither the light source nor the observer is at infinity with respect to the lens, a correction factor needs to be applied, which has the form
\be 
\bar{\alpha} = \frac{D_{\rm LS}}{D_{\rm S}} \alpha, \label{eq:alpha_bar}
\ee
where $D_{\rm LS}$ is the lens-source, and $D_{\rm S}$ the observer-source distance respectively. We will further denote the observer-lens distance with $D_L$. The ratio $\frac{D_{\rm LS}}{D_{\rm S}}$ is essentially the ratio of the size of the binary system over the distance to the observer. To get an idea, if the distance of the star from the lens is $D_{\rm LS}\sim 1.4AU \approx 300 R_\odot$, while the distance of the system from the observer is $D_{\rm S}=480 pc\approx 2.127\times10^{10}R_\odot\approx D_{\rm L}$, this implies that $D_{\rm LS}/D_{\rm S}\approx 1.4\times10^{-8}$. The numbers here are inspired by the BH1 system \citep{El-Badry2023} and in our discussion we will often use the parameters of this system as a template for estimates. 

\section{Dark halo light curve (no central black hole)} \label{sec:magnification}

Here, we will solve the lens equation to compute the structure of the multiple images and, subsequently, the resulting magnification and light curve, assuming an isolated DH without a BH at its centre. As has been previously argued, these observables could possibly be the smoking gun signatures for a DH in photometric observations, given the relevant systematics. The lensing equation reads 
\be 
\beta \equiv \theta - \bar{\alpha}(\theta), 
\ee
where $\theta$ is the apparent position of the light source in the observer's sky, $\beta$ is the true position, and $\bar{\alpha}$ is defined by equations \eqref{eq:deflection} and \eqref{eq:alpha_bar}. We express the impact parameter in terms of its angular size, $b = D_L \theta$, and the angular size of the lens as $\theta_R=R/D_L$, where $D_L$ is the observer-lens distance. The full lens equation then reads as
\be u = x - \frac{1}{x} \left\{   
          \begin{tabular}{cc} 
           $1$ & $,x\geq x_R$\\  
           $ 1-\left[1 - (x/x_{R})^2\right]^{3/2}  $& $,x< x_R$
           \end{tabular}.\right. \label{eq:u-x}
\ee
In this equation, we defined the angle $\theta_{\rm E}^2 \equiv \frac{D_{LS}}{D_S D_L}\frac{4GM}{c^2}$ as the angular size of the standard Einstein ring (i.e., in the case of a point source), and normalised the rest of the angles as $u \equiv \beta/\theta_{\rm E}$, $x \equiv \theta/\theta_{\rm E}$, and $x_R \equiv \theta_R/\theta_{\rm E}$. Clearly, for $\theta > \theta_R$ the lens behaves as a point mass, while for $\theta \leq \theta_R$ we start to see the effects of the DH. 
To give a sense of scale, for a binary with a lens mass of $\sim10M_\odot$ and an orbit of size $\sim300R_\odot$, such as in the case of {\it Gaia} BH1 \citep{El-Badry2023}, a DH of a radius equal to $1R_\odot$ will have $x_R\sim6$.

The lens equation can be solved numerically to determine the images that are produced by the DH, as roots of the equation
\be 
x-u =\tilde{\alpha}(x),  \label{eq:lens-equation}
\ee
where we have redefined $\tilde{\alpha}=\bar{\alpha}/\theta_E$.
This essentially corresponds to looking for the $x$-value of the intersections of the lines $y=x-u$ (l.h.s) with the curve $y=\tilde{\alpha}(x)$ (r.h.s) for different values of the normalised true position of the luminous source (i.e., $u$). 

 \begin{figure}
 \centering
\includegraphics[width=0.4\textwidth]{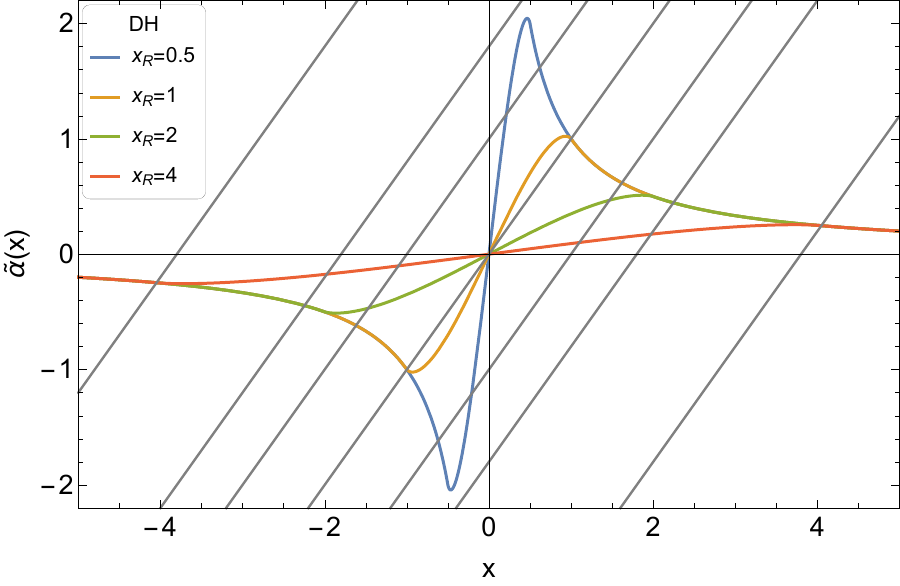}
\caption{Graphic solution of the lens equation \eqref{eq:lens-equation} for different DHs, each corresponding to different sizes parametrised by the quantity $x_R$ (see equation \eqref{eq:u-x}). Here, we assume that the DH has a constant density, as well as the whole of the dark object is composed of a DH and that no point-like mass such as a BH exists at its centre. The gray lines correspond to the l.h.s of equation \eqref{eq:lens-equation} and their intersection with each of the coloured curves (the r.h.s of \eqref{eq:lens-equation}) yields a solution for a multiple image. These curves give essentially the deflection angle of a light-ray.}
\label{fig:DHlenseq}
\end{figure}

This process is shown in Figure~\ref{fig:DHlenseq} for DHs of various sizes. As can be seen, for every value of $u$ there exist either one or three intersections between the lines of $y=x-u$ and the respective curves of $y=\tilde{\alpha}(x)$, which correspond to one or three images of the light source created by the DH lens. It is worth noticing that varying $u$ corresponds to evolving the binary system in time. One can show that the slope of $\tilde{\alpha}$ at $x=0$ is $\left.\frac{d\tilde{\alpha}}{dx}\right|_{x=0}=\frac{3}{2x_R^2}$, which means that when the size of the lens is $x_R\geq \sqrt{3/2}$ the slope is less than $1$ and there can be a maximum of one image since there can be only one intersection between $y=x-u$ and $y=\tilde{\alpha}(x)$, while when $x_R< \sqrt{3/2}$ there can be as many as $3$  \citep[see also][for discussion on this]{Croon_2020a}. In general, when the true position of the light source is far from the axis connecting the observer to the lens (i.e., large values of $u$) there is only one image produced by the DH lens from the intersection of $y=x-u$ with $y=1/x$. We remind here that, in this case, we consider a DH without a point-like mass (BH) at its centre. We will get back to this issue later. 

We mention here that for the more compact DH lenses with $x_R<1$, when the light source passes through $u=0$ one can observe both an Einstein ring at $\theta_E$ as well as an image that will look like a point source at $\theta=0$ (see in Figure~\ref{fig:DHlenseq} the line $y=x$). The ring and the point source correspond to the total of three images that one in general gets for the more compact sources with $x_R<\sqrt{3/2}$, as shown also in the examples of Figure~\ref{fig:DHlenseq}.
The ring forms just as in the case of a point mass when the light source is exactly behind the lens. In that case, the corresponding images get deformed in the direction perpendicular to the radial on the image plane and form a circular ring.

The total magnification is the sum of the individual ones,
\be\label{eq:mu_tot} \mu_{\rm tot}=\sum_i |\mu_i|,\ee 
with 
\be \label{eq:mu_def1}
\mu_i = \frac{\theta}{\beta}\frac{d\theta}{d\beta}  = \frac{x}{u}\frac{d x}{d u},
\ee 
where the derivative $d x/ d u$ can be easily computed through the lens equation as
\be \label{eq:mu_def2}
\frac{dx}{du} = \left( 1 - %
\frac{d \tilde{\alpha}(x)}{dx} \right)^{-1}.
\ee
These DH lenses have two key phenomenological effects when they are compact enough. In particular, when the light source starts to pass behind the lens, there is a strong magnification effect for one of the images that affects the appearance of the source. This is due to the fact that there is an image solution at position $x$ for which $\frac{d\tilde{\alpha}(x)}{dx}=1$ that causes $\frac{dx}{du}$ to diverge. The other effect is that we have a magnification divergence at $u = 0$ for the images which lie at a normalised angle with $x \neq 0$. 

%
 \begin{figure}
 \centering
\includegraphics[width=0.4\textwidth]{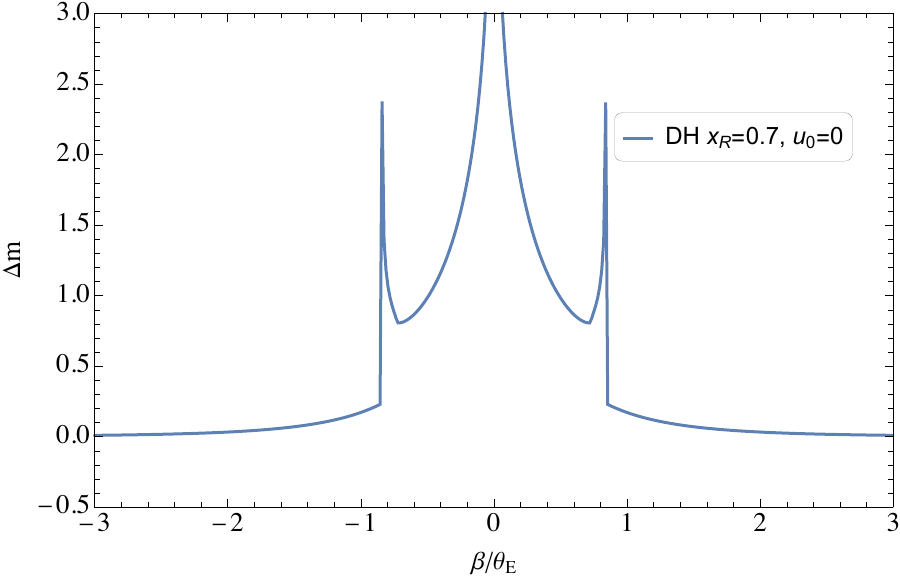}
\caption{The change in magnitude $\Delta m$ of a light source during an eclipse by a DH lens of size $x_R=0.7$, as a function of the true position of the source on the sky, the latter parametrised by the angle $\beta$ (see equation \eqref{eq:Deltamu}). We assume that the lensing potential is solely produced by the DH, i.e., no BH exists at the centre of the DH. We remind that $\Delta m$ is directly related to the total magnification $\mu$, defined in equation \eqref{eq:Deltamu}. Our results appear to be in qualitative agreement with \citet{Romao_2024}.}
\label{fig:mag2} 
\end{figure}
%

Finally, from the magnification, one can estimate the change in magnitude of the light source as it passes behind the lens for different values of the normalised true position $u$ according to 
\be \Delta m = 2.5\log_{10}\mu_{\rm tot}. \label{eq:Deltamu} \ee
This is shown in Figure \ref{fig:mag2} for a DH of size $x_R=0.7$. We can see in this figure the spikes in magnification due to the two effects we discussed above (our sign convention in Eq. \eqref{eq:Deltamu} gives positive $\Delta m$ when the source gets brighter).

The time scale characterising the duration of the lensing event is the Einstein time $t_E$, 
\be
t_E = \frac{D_L \theta_E}{v_{\perp}}, \label{eq:t_E}
\ee
where $v_{\perp}$ corresponds to the relative velocity of the source-lens in the direction perpendicular to the line-of-sight of the observer (and in our case is related to the orbital velocity of the star). The presence of a DH would affect $\theta_E$, and possibly the velocity $v_{\perp}$ if the trajectory of the source is modified due to the difference in the gravitational potential and/or the existence of dynamical friction effects. We will not explore here such effects, and leave them for future research. It is important to notice that the physical quantities which enter into the Einstein time cannot be measured by lensing alone, and more information is needed. Astrometry observations can indeed break the degeneracy between these parameters through a measurement of the velocity. To model the actual light curve associated with the lensing effect in the time-domain we assume a trajectory for the source as 
\be
u(t)^2 = u_{0}^2 + \left( \frac{t-t_0}{t_E} \right)^2, \label{eq:u(t)}
\ee
with the impact parameter $u(t_0) = u_{0}$, being the closest approach between lens-source. We can understand the scaling of the luminosity curve by plugging equation \eqref{eq:u(t)} into the magnification \eqref{eq:mu_tot}.

Assuming a Keplerian orbit for the light source around the DH lens (i.e., $v_\perp\approx \sqrt{\frac{GM}{A}}$, with $A$ the semi-major axis of the orbit which is $A\approx D_{\rm LS}$) and under our main assumption that $D_{\rm L}\approx D_{\rm S}\gg D_{\rm LS}$, the Einstein time is $t_E = 2D_{\rm LS}/c$. Using Kepler's laws the Einstein time can be expressed as a function of the period $T$ and the total mass of the system $M$ as 
\begin{equation}
t_E = \frac{2}{c} \left( \frac{T}{2 \pi} \right)^{2/3} \left( G M \right)^{1/3}.
\end{equation}
If we express the mass in $M_\odot$ and the period in years, then this becomes 
\be t_E\simeq 16.6 ~{\rm min} \left(\frac{M}{M_\odot}\right)^{1/3}\left(\frac{T}{\rm yr}\right)^{2/3}.
\label{eq:t_EinMin}
\ee

\begin{figure}
 \centering
\includegraphics[width=0.4\textwidth]{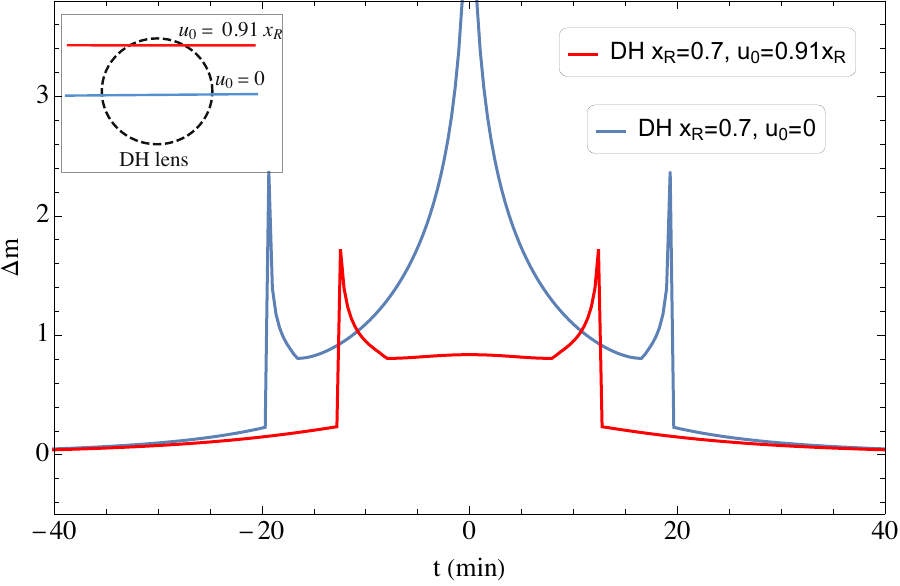}
\caption{Similar $\Delta m$ to Figure \ref{fig:mag2}, but now as a function of time. The evolution depends on the characteristic time $t_E$ (see \eqref{eq:t_E}), and we have assumed the same characteristics as those of the {\it Gaia} BH1 system. The figure shows two transit events, one with initial impact parameters $u_0=0$ and $u_0=0.91x_R$, respectively. The transit for the first event lasts a total of $\sim40$min, while the spikes in the magnification that happen as the source passes behind the edge of the DH lens last $\sim5$min each. For the rest of the transit, the source is brighter than its normal brightness and behaves like a point mass. The second transit lasts less time, $\sim25$min since the light source grazes the lens. In this case, we observe two magnification spikes as the source goes behind the lens, while the source remains magnified throughout the transit. This plot should be compared with similar ones for e.g. by \citet{Romao_2024}.}
\label{fig:mag4}
\end{figure}

For a system like the {\it Gaia} BH1, the total mass is $\sim11M_\odot$, the mass ratio between the total mass and the luminous-star mass is $\sim14$, the period is $185.6$~d, and the size of the orbit is $\sim 1.4AU$ \citep{El-Badry2023}. For such a system the Einstein time is of the order of $\sim 23$ min. The time evolution of the apparent magnitude of the source as it passes behind the lens can be seen in Figure~\ref{fig:mag4} for the case of two occultation events, the first with $u_0=0$ and the second with $u_0=0.91x_R$, respectively. Although the orientation of the BH1 system is not favorable for such an observation, other similar systems may be optimally oriented.

\section{Dark halo light curve (Black hole with a dark halo)} 
\label{sec:magnification2}
We can extend our previous discussion of a DH of uniform density by including a point mass at the centre of the halo. This case would correspond to the situation where a compact object such as a BH has formed a dark matter halo around it, or developed a configuration of exotic particles of bosonic nature (hair). Our discussion is generic and can be applied to BHs of different masses and DHs of different nature, respectively.

The lens equation will be as before with a correction due to the fact that we have two mass sources, the point mass at the centre and the DH, i.e.,

\be u = x - \frac{1}{x} \left\{ 
          \begin{tabular}{cc} 
           $1$ & $,x\geq x_R$\\  
           $ 1-q^{-1}\left[1 - (x/x_{R})^2\right]^{3/2}  $& $,x< x_R$
           \end{tabular},\right. \label{eq:lens-eq-BH-DH}
\ee
where $q=M_{\rm tot}/M_{\rm DH}$ is the ratio of the total mass of the lens (BH+DH) over the mass of the DH, while the rest of the quantities are defined as before.

 \begin{figure}
 \centering
\includegraphics[width=0.4\textwidth]{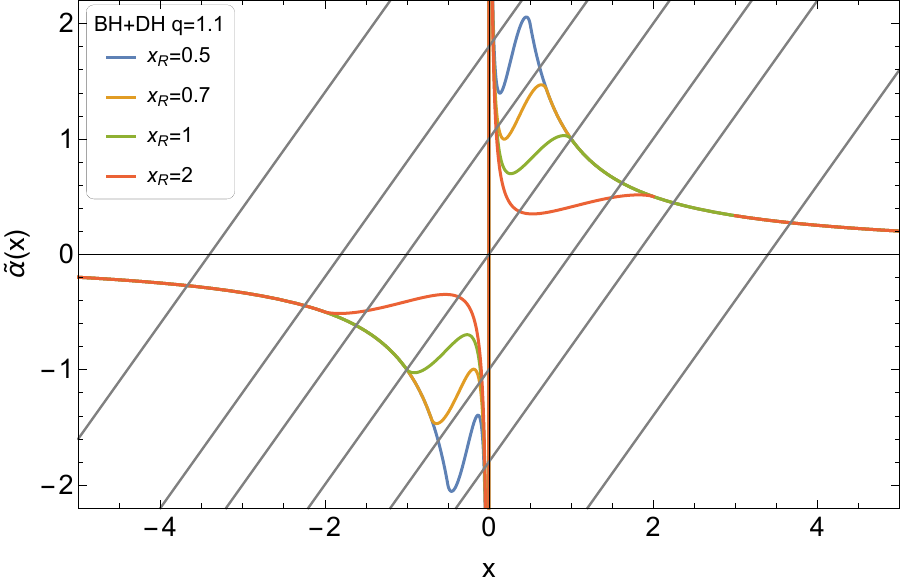}
\includegraphics[width=0.4\textwidth]{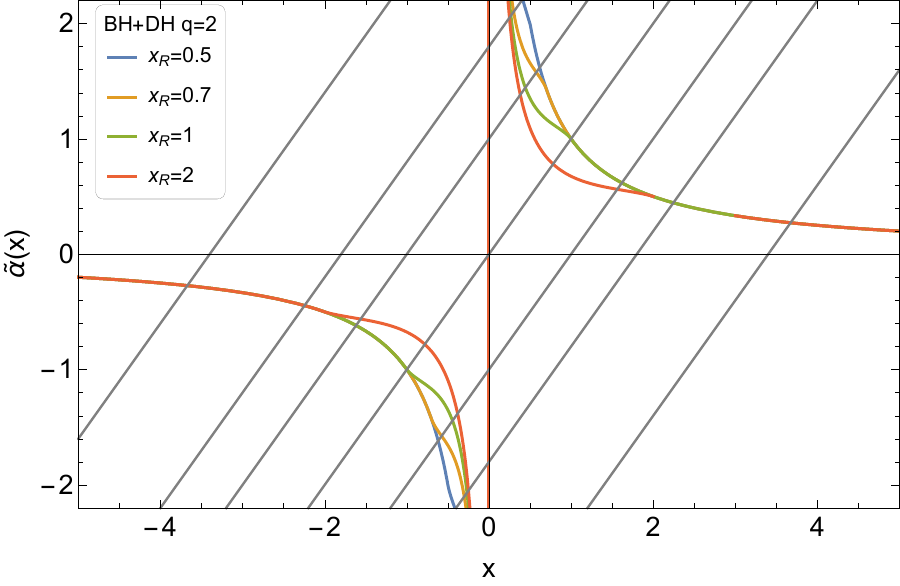}
\caption{
Same as Figure \ref{fig:DHlenseq}, but for the case of a DH with a point-like mass at its centre (see equation \eqref{eq:lens-eq-BH-DH}). This corresponds to the case where the exotic matter has clumped around a BH. The mass ratio $q$ is defined as $q=M_{\rm tot}/M_{\rm DH}$, where $M_{\rm tot}$ the total mass of the lens (BH + DH). The different colours correspond to the graphical solution of the lens equation for DHs of different sizes $x_R$. In this case, the curves for $\tilde{\alpha}(x)$ diverge at $x=0$ due to the presence of the point mass at the centre.
}
\label{fig:BHDHlenseq} 
\end{figure}

As in the case of a pure DH, one can solve the lens equation thus obtaining the images for different values of $u$, as can be seen in Figure~\ref{fig:BHDHlenseq}. One can see that the qualitative behaviour for the BH+DH system is a little different. The main difference is that since there is a point mass at the centre, there is a divergence at $x=0$, where the point mass dominates. Between the points $x=0$ and $x=x_R$, there is a region where the DH contribution becomes significant too, while as we approach $x_R$ we have a surface effect which is similar to the one occurring in the case without a central BH. For values of $q$ close to $1$ there can be more images than in the DH-only case due to the turnover that the function $\tilde{a}(x)$ makes (see also Figure \ref{fig:BHDHlenseq}). We should also point out that when $q$ becomes large, i.e., the mass of the DH is much smaller than the total mass, then the lens behaves as a pure point mass lens. The lens behaves as a point-mass-lens also in the case of very compact lenses ($x_R<1$) when the light-source is behind the lens, i.e., for $u$ in a region around zero.

\begin{figure}
 \centering
\includegraphics[width=0.4\textwidth]{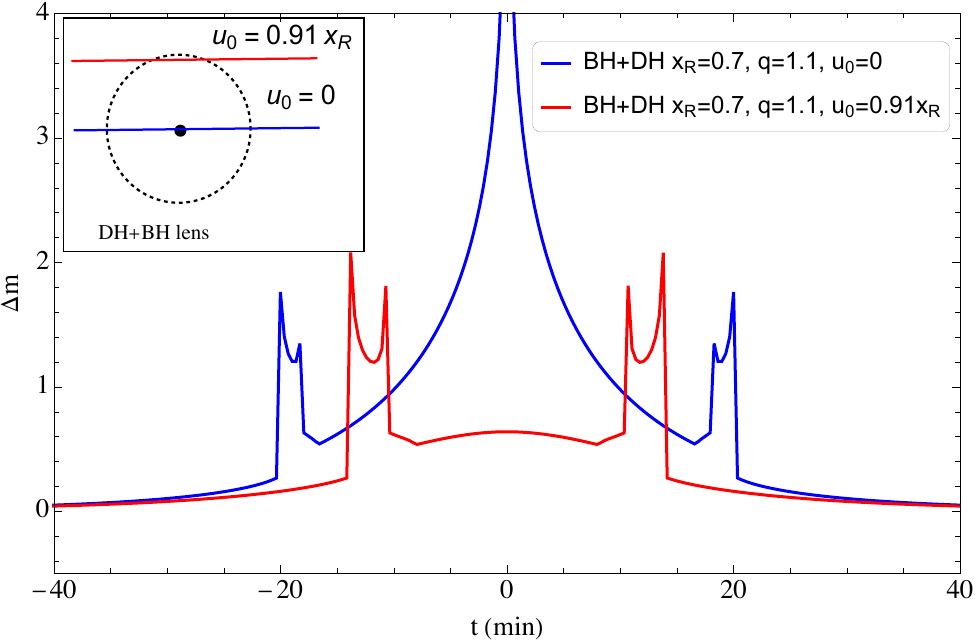}
\caption{Time evolution of the apparent magnitude change $\Delta m$ of a light source during an eclipse by a BH, surrounded by a DH (BH+DH). We assume the size of the halo to be $x_R=0.7$, and the mass ratio $q \equiv M_{\rm tot}/M_{\rm DH} =1.1$. We remind that $u_0$ denotes the initial impact parameter. This case is to be contrasted with Figure \ref{fig:mag4}, where there is no point-like mass at the centre of the DH.}
\label{fig:mag3}
\end{figure}

One can then estimate the magnification for this lens in the same way that we did for the DH and from that, the change in apparent magnitude. The resulting change in magnitude of a light source that passes behind the lens can be seen in Figure~\ref{fig:mag3}. For the case of the BH+DH, the magnification behaviour is similar to the DH-only case with the difference that the turnover of $\tilde{a}(x)$ presents more points where $\frac{d\tilde{\alpha}(x)}{dx}=1$ and therefore more instances for which the magnification can diverge. This can be seen in Figure~\ref{fig:mag3} as the double-peaked features, which should be compared with the single-peak features of the DH-only case shown in Figures~\ref{fig:mag2} and \ref{fig:mag4} respectively.

\section{Detection probability with the Gaia sampling}
\label{sec:Detection}
Gaia samples the sky in a semi-regular and predefined manner, following the Nominal Scanning Law. There were two exceptions to this law, the first one at the beginning of the mission, the spacecraft rotation axis was maintained in the ecliptic plane, the second involving a change in the precession rotation axis during one year. Observation times can be obtained through the \textsc{GOST} (Gaia Observation Forecast Tool) web interface. We use these observation times to determine the probability of observing an event as described in the above sections.

The evaluation of the detection probability is derived from the following hypothesis:
We selected 10,000 directions uniformly distributed on the sphere, assuming that the source might preferably be in the halo. A periodic event is chosen to reoccurring with the period $P$ (days), the length of the event is short, of duration $d$ (minutes). We determine if this event is covered by an observation in one of the 10,000 directions.

A Gaia Field-of-View epoch consists of a series of independent photometric measurements (lasting about 40 seconds), including the Sky Mapper, nine consecutive CCD measurements of the Astrometric Field in the $G$ band, as well as measurements of the integrated spectrophotometry $G_{BP}$, $G_{RP}$, and, if the source is sufficiently bright, of the Radial Velocity Spectrometer, the $G_{RVS}$ (in the near-infrared). The lensing effect is significant (greater than 0.1 mag) and unambiguously detectable at Gaia's precision level. Since these independent measurements occur over a very short period, they allow us
to describe short-timescale variations in the event and help eliminate single outliers, thereby reducing false detections.

\begin{figure}
 \centering
\includegraphics[width=0.45\textwidth]{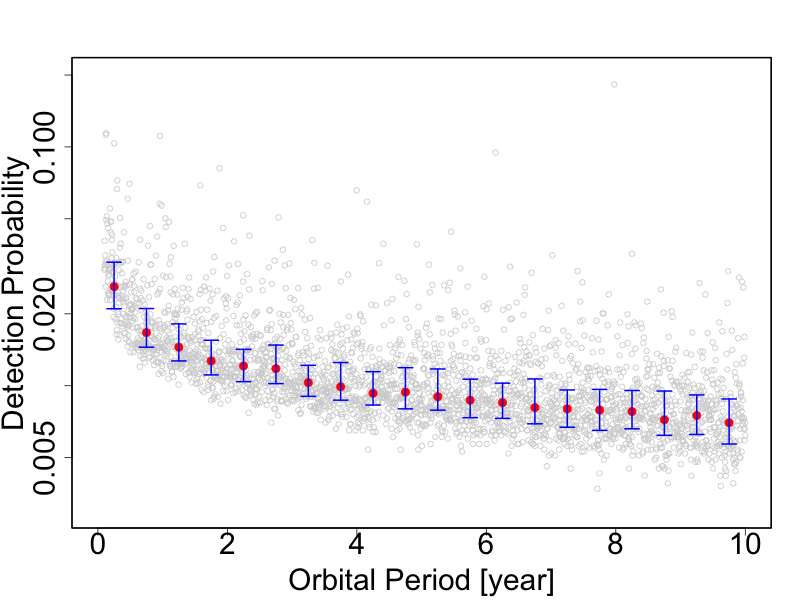}
\caption{ Detection probability of a lensing effect from a 10\,$M_\odot$ BH as a function of the orbital period with Gaia. The plot shows the median value, along with the first and third quartiles.}
\label{fig:detectionProb} 
\end{figure}

In Fig.~\ref{fig:detectionProb}, we show the simulation result for a 10 solar mass BH. The duration of the event is taken from Eq.~\eqref{eq:t_EinMin}.
%
%

In the simulation, the probability of detecting a single event with a 10-year period is 1 in 130. The result becomes more robust when multiple detections occur, so shorter orbital periods have an advantage. Additionally, if the BH's mass is greater, the probability of detection increases because the events last longer.

There are two main approaches to detecting such systems with Gaia. The first is to analyse in detail the photometry of astrometric and radial velocity published BHs and assess whether short-timescale variations is being detected in the per-CCD data. The second is to conduct a general search for repeated variations in the per-CCD data within the Field of View (FoV) transits. This is feasible since all sources will be published along with their per-CCD data in DR4. Even though this is an enormous task with the possibility of many false positives, the knowledge of the expected times of the lensing events makes it more tractable. 
In both cases, ground-based follow-up observations can be planned if the period is well constrained and not too long. This is, therefore, a strong incentive to search for such objects.

\section{Summary and discussion} \label{sec:summary}
Using the established microlensing framework in the literature, we have discussed a probe of dark matter and other exotic configurations in our galaxy based on the synergy between astrometry and weak lensing. The recent discovery of the binary systems BH1, BH2, and BH3 by the {\it Gaia} mission is only the beginning of an exciting and unique opportunity to probe fundamental physics with the high precision astrometry and photometry frontier. Here, we have demonstrated that the photometry of such binary systems allows for the indirect probe of exotic configurations in our galaxy through their weak lensing effect, ultimately imprinted on the resulting light curve. As explained around equation \eqref{eq:u-x}, the modification of the gravitational potential, due to the existence of a DH, drastically changes the structure of multiple images and their magnification, compared to the standard case of a central potential. This holds for both the case of an isolated DH (see Figures \ref{fig:mag2}, \ref{fig:mag4}), as well as the case of a DH surrounding a central BH (see Figures \ref{fig:BHDHlenseq}, \ref{fig:mag3}). Our results are in qualitative agreement with more general ones derived in the literature, as discussed in the text. Our aim was not to present a new formalism, but rather apply it to {\it Gaia}-like binaries.

Our results suggest that precision photometric observations of galactic binaries accompanied by astrometric observations, can detect and/or constraint exotic configurations such as dark matter halos. 

 \begin{figure}
 \centering
\includegraphics[width=0.45\textwidth]{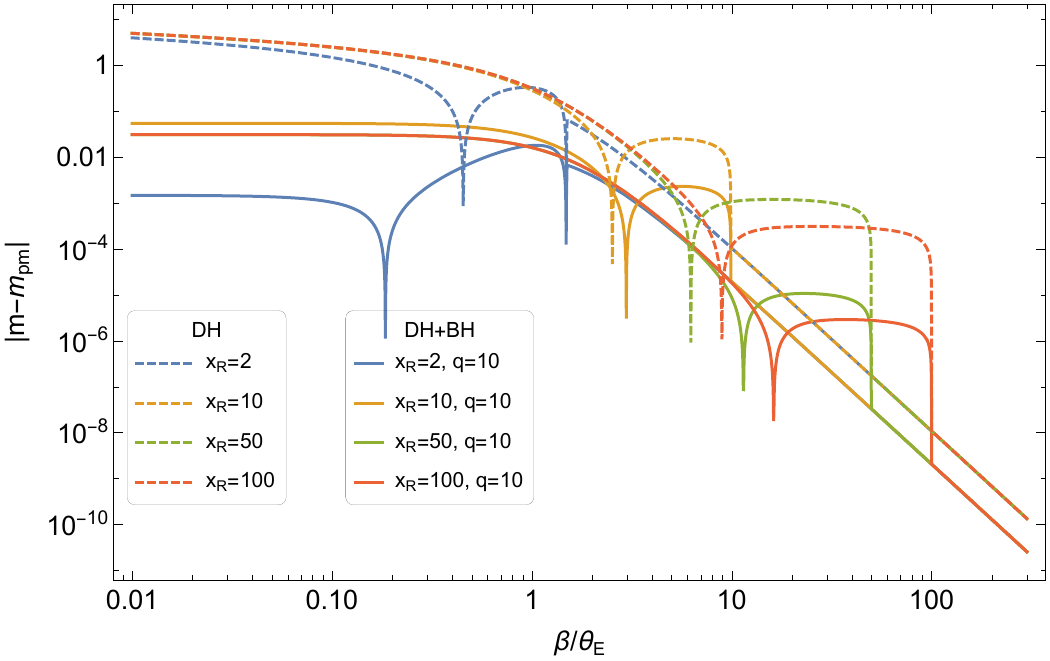}
\caption{Plots of the absolute difference between the expected magnitude due to lensing by a point mass lens, $m_{pm}$, and the magnitude $m$ due to lensing by a DH (dashed) and a DH+BH (solid) system of the same mass. The spikes in the curves are due to the fact that the $y$-axis is logarithmic and there are some values of $\beta/\theta_E$ for which $|m-m_{pm}|=0$. 
The magnification $\mu$ for the DH or the BH+DH far from the lens is always a little smaller than that of the point mass. Then, near the surface of the DH or the DH+BH, $d\tilde{\alpha}/dx$ becomes a little less steep than what it is for the point mass lens (see Figures \ref{fig:DHlenseq} and \ref{fig:BHDHlenseq}), raising $dx/du$ closer to the value of 1 (see eqs. \ref{eq:mu_def1},~\ref{eq:mu_def2}) thus making $\mu$ larger from that of a point mass, which introduces the first outermost crossing near $\beta/\theta_E=x_R$. Then further in, the point mass $\mu$ becomes larger again, as explained in the text, therefore causing the second crossing.}
\label{fig:magDif}
\end{figure}

At this point, we should emphasise an important difference between isolated DHs sought-for in surveys, thus far \citep{Alcock_2000,EROS-2:2006ryy,2011MNRAS.416.2949W,2024Natur.632..749M}, and the DHs that are parts of binary systems expected in {\it Gaia}'s data. As pointed out by \cite{Croon_2020a}, the lenses where one can have a detectable signal in the former case, are those with a size comparable to $\theta_E$ (i.e., $x_R\approx 1$, see Figures \ref{fig:mag4} and \ref{fig:mag3}). In the latter case though, one can have access to less compact systems ($x_R\gg 1$). {\it Gaia}'s astrometric observations can measure the masses of the binary members, as well as their orbits (size and inclination) within some observational precision. Knowing the mass of the lens and the geometry of the lens-light source system can allow for measurements of weaker signals, i.e., the difference between the expected brightening for a point mass lens from the observed brightening due to a DH, $|m-m_{pm}|$, for a given mass (see Figure \ref{fig:magDif}). 
One can see that in the case of a DH-only lens, there is a difference from the expected brightness from a point-mass-lens for $u\rightarrow 0$, which is due to the diffuse DH mass providing a lot less magnification, while for a DH+BH lens, there is a difference because the BH at the centre of the DH, lenses the light source as a point-mass-lens but with less mass than the total mass of the DH+BH system as estimated by the orbital dynamics inferred by the astrometry. One could say that there will be a mass discrepancy between the astrometry and the weak lensing in the latter case. Thus {\it Gaia} systems will give access to DH systems not available to previous surveys. 

This, however, is the first step, relying on simplifying assumptions, such as a DH of constant density, an ideal orientation of the binary's plane with respect to the observer and a point size source. When confronted with observations, these assumptions must be challenged. 
For example, the finite size of the light source, i.e., the companion star, is something that needs to be taken into account in order to have a realistic model for the light curve from these binaries \cite{Witt1994,Matsunaga:2006uc}. The finite size of the light source is expected to smooth out the sharp features of the magnification profiles for the DH systems, as it does for the point mass. However, depending on the situation, either the total magnification or the shape of the light curve, or both, are expected to be different, in particular due to the features that exist near the outer parts of the DH distributions. These features are, for example, expected to modify the light curve during the start and the end of a transit, while the extended nature of the DH will affect the light curves in a way distinctive from that of a point mass even in the case when the center of the lens is well outside the apparent disk of the star. All these effects will be explored in detail in a forthcoming investigation.
Furthermore, for an analysis with additional mass profiles for the DH we refer the reader to \cite{Romao_2024}.

Before closing, it is interesting to briefly discuss the expected number of galactic binary systems such as BH1 or BH2 which {\it Gaia} could detect. According to the publicly available catalogue by \cite{Olejak:2019pln}, based on population synthesis methods with the {\it StarTrack} code, for the case of BH - main sequence star (giant star) binaries, simulations predict that there are about $10^5$ ($7\cdot 10^3$) of them in the thin disk, $8\cdot 10^3$ ($2\cdot 10^2$) in the halo, and $10^4$ ($3\cdot 10^2$) in the thick disk of the galaxy respectively. The orbital periods of the binary systems $BH1$ and $BH2$ discovered by {\it Gaia} DR3, suggest that the distribution of periods is bimodal with a flat part in between 1 and 3 years. The upcoming releases, DR4 and DR5 are expected to discover a few tenths of such systems due to the longer observing window and the less stringent restrictions on the signal-to-noise ratio compared to the {\it Gaia} DR3 \citep{El-Badry2023}. How many of these systems will be favourable for the type of observations we have discussed or estimating the probability of finding such systems, is beyond the scope of this work and will be the subject of future investigation.

In conclusion, we need to emphasise an important aspect of binary systems with DHs and stelar companions. The observation of specific, astrometrically-resolved binaries is fundamentally different to previous probes which focused on searching for serendipitous, one-time events. The {\it Gaia} DH binary systems will be available to observe repeatedly as the star orbits the DH lens (in addition, we will be able to anticipate the occurrence of these lensing events with relative accuracy), thus allowing for higher precision measurements of the structure of the DH.  

\section*{Acknowledgments}
We are indebted to Benoit Famaey for invaluable suggestions, and thank Bradley Kavanagh for discussions. We also thank Lukasz Wyrzykowski for very useful discussions and his feedback. We also thank the referee for their useful feedback.




\section*{Appendix: BH and DH system}
Here we derive the gravitational potential and the deflection angles for the case of a uniform solid sphere (i.e sphere with constant density), with a point mass at its centre, the latter representing the central BH which may exist at the centre of the DH. We use the same notations as in the text, but we set $c = 1$ for simplicity. 

We work within the Newtonian limit, and start from the Poisson equation for the gravitational potential $\Phi$ and within the spherical DH (i.e $r \leq R$),
\begin{equation}
\nabla^2 \Phi(r) \equiv \frac{1}{r} \frac{d^2}{dr^2} \left( r \Phi(r) \right) = 4 \pi G \rho,
\end{equation}
where $\rho = M_{\rm DH}/\left(\frac{4}{3} \pi R^3 \right)$ corresponds to the constant density of the DH. We can easily solve the above equation as
\begin{equation}
\Phi(r \leq R) = \frac{2 \pi G \rho}{3} r^2 + \frac{K}{r} + c,
\end{equation}
with $K,c$ constants of integration to be determined from appropriate boundary conditions. As $r \to 0$, we expect the potential to be dominated by the central potential of the point mass (BH), therefore, $\Phi(r \to 0) = - G M_{\rm BH}/r$, which fixes the constant $K$. The constant $c$ is determined by requiring continuity of the potential at the surface of the halo, $r = R$. It is, however, irrelevant for our analysis as this constant will be eliminated by the $\partial/\partial b$ derivative in the expression for the deflection angle. Outside the halo we have $\rho = 0$ and the potential is
\begin{equation}
\Phi(r \geq R) = - \frac{G M_{\rm tot}}{r},
\end{equation}
with $M_{\rm tot} \equiv M_{\rm DH} + M_{\rm BH}$. Evaluation of the equation \eqref{eq:deflection} then leads to 
\begin{align}
\alpha & = \frac{4 G M_{\rm tot}}{b} + \frac{4G M_{\rm DH}}{b R^2}  \cdot a \cdot  \left( \frac{ b^2 }{R} - R \right) \nonumber \\
& = \frac{4 G M_{\rm BH}}{b}  + \frac{4 G M_{\rm DH}}{b} \cdot \left( \frac{b^2 a}{R^3} - \frac{a}{R} + 1 \right) \nonumber \\
& = \frac{4 G M_{\rm BH}}{b}  + \frac{4 G M_{\rm DH}}{b} \cdot \left(1- \frac{a^3}{R^3}\right) \nonumber \\
& \equiv \alpha_{\rm BH} + \alpha_{\rm DH}
\end{align}
reminding that $a^2 = R^2 - b^2$ and we set $c = 1$. Therefore, the deflection angle for the general case of a BH+DH is the sum of the respective angles for an isolated BH and DH.



\bibliography{example} 

\end{document}